# Low-magnitude Seismicity with a Downhole Distributed Acoustic Sensing Array – examples from the FORGE Geothermal Experiment

**A. Lellouch[1], R. Schultz[1], N.J. Lindsey[1], B.L. Biondi[1], and W.L. Ellsworth[1]**

[1]Department of Geophysics, Stanford University, Stanford, California, USA

Corresponding author: Ariel Lellouch (ariellel@stanford.edu)

**Key Points:**

- A downhole Distributed Acoustic Sensing (DAS) array is used to study seismicity around the FORGE Enhanced Geothermal System experiment.
- Within a 10.5-days period, 82 earthquakes are detected using the DAS array, compared to four in the regional catalog.
- Detected events originate from historically active source areas and can be clustered into distinct families.

**Abstract**

We show the capabilities of a downhole Distributed Acoustic Sensing (DAS) array in detecting, locating and characterizing low-magnitude earthquakes occurring in the vicinity of the Frontier Observatory for Research in Geothermal Energy (FORGE) site in Utah. 10.5 days of continuous data were acquired in a monitoring well at the FORGE geothermal site during the initial stimulation of an Enhanced Geothermal System in April-May 2019. Earthquake activity beneath Mineral Mountains, Utah also occurred within 10 km of the FORGE monitoring well. During the experiment, four events from those areas were cataloged by the University of Utah Seismograph Stations. Our processing of DAS data, including template matching, finds 82 earthquakes during that period, of which 16 are visible on the regional network. The magnitude of completeness obtained by DAS processing is better by at least M=0.5 than the dense surface array around the FORGE site. While a single vertical DAS array is limited in terms of event location due to its azimuthal ambiguity, multiple DAS wells or a combination of a downhole array with surface stations or near-surface horizontal DAS could jointly resolve locations. All detected events probably originated from the two active source areas, located 3-5 and 8-10 km away from the FORGE site, respectively. Recorded events can be clustered into several distinct families.

**1 Introduction**

Enhanced Geothermal Systems (EGS) have great potential for baseload low-carbon energy. However, in order to be economically feasible, the hot rock mass has to be sufficiently permeable to allow for economical amounts of fluid to pass through it, accumulating heat in the process (Tester et al., 2006; Majer et al., 2007). In order to increase permeability, techniques often used in tight unconventional hydrocarbon extraction are employed. In hydrofracturing, the fluid pressure in the well is increased until it exceeds the fracture gradient of the rock formation. As a result, a tensile failure occurs, and fluid permeates the newly opened mode-I crack. The fracture keeps propagating away from the well until the pressure drops below the fracture gradient. However, shear failure, or mode II, has also been associated with hydraulic fracturing (Martínez-Garzón et al., 2013; McClure and Horne, 2014). Microseismic monitoring is the primary tool to detect, locate and characterize weak seismic waves associated with fracture opening (Maxwell, 2014). Through the use of proppant, fractures can be prevented from closing once pressure in the well is returned to normal. As such, fluid can be circulated through them with much lower pressure. For an EGS project to be successful, the created fracture network has to connect an injection and production well with sufficient flow between the two without short-circuiting due to preexisting or newly created flow paths. While EGS experiments have been conducted for more than three decades (Fehler, 1989; Garcia et al., 2012), understanding fracture networks and their behavior has remained elusive, thus limiting the economic success of EGS.

To address those limitations and others, the U.S. Department of Energy created the Frontier Observatory for Research in Geothermal Energy (FORGE) experiment, a dedicated underground field laboratory in Utah whose purpose is to develop, test, and accelerate breakthroughs in EGS. It is located near the town of Milford in Beaver County, Utah, on the western flank of the Mineral Mountains (Figure 1), close to the active Blundell geothermal plant in the Roosevelt Hot Springs (Ross et al., 1982; Gwynn et al., 2016). Phase 2-C of the experiment, conducted in 2019 between April 21$_{st}$ to May 3$_{rd}$, consisted of hydraulic stimulation of the target rock and recording seismicity in a nearby monitoring well, shallow boreholes, and on the surface. Various stimulation regimes were tested to find the optimal method for fracture network creation. The resulting seismic activity can be coarsely separated into two categories – microseismic and induced seismicity. Microseismic events are caused by the energy released when fractures open due to stimulation and are generally considered to have negative magnitudes. Induced seismicity, on the other hand, refers to earthquakes caused by stimulation and/or fluid injection (Lee et al., 2019; McGarr et al., 2015; Schultz et al., 2020). As several EGS projects have been previously halted due to induced seismicity (Deichmann et al., 2009; Grigoli et al., 2018; Ellsworth et al., 2019), it is important to monitor the seismic activity.

Downhole geophones are highly sensitive tools for seismic monitoring (Maxwell et al., 2012). They offer three axis of measurement, high sensitivity, and have well-known response functions. Through polarization analysis, they can provide information on the events' direction of arrival. However, they do have several downsides. First, the operating temperature and pressure for conventional digital tools is limited. In EGS conditions, where the target rock reaches temperatures of more than $200_0$ C, deploying conventional geophones close to the reservoir is impractical, as these sensors are not designed for long-term monitoring in such areas (Zhidong et al., 2019). There are other designs that avoid deploying electronic circuits in the high-temperature zones, notably fiber-based sensors, but they require deploying additional wires in the well. In addition, all downhole geophones prohibit almost any operation in the well in which they are installed. As a result, a dedicated monitoring well has to be drilled, which incurs large additional costs. Any deployment or retrieval of the geophones in the well also requires significant time and effort. In practice, downhole geophones are often only deployed for microseismic monitoring during stimulation, whereas surface arrays are used to monitor induced seismicity continuously. Since potential induced events of concern are larger than stimulation events, a surface array can be expected to detect them despite higher anthropogenic noise, near-surface scattering, and anelastic losses due to propagation in sediments. The FORGE experiment followed that approach, and downhole recording is available only for the stimulation period, while a surface array was permanently installed.

An enticing alternative for both short- and long-term monitoring of EGS is the use of Distributed Acoustic Sensing (DAS). In DAS, an optical fiber is turned into a seismic sensor thanks to a dedicated optical apparatus, known as an interrogator, that continuously sends laser pulses through the fiber. DAS has been deployed in wells for almost a decade, mostly in the oil and gas industry (Jin and Roy, 2017; Karrenbach et al., 2019; Mateeva et al., 2014, 2013) but also to study tectonic seismicity (Lellouch et al., 2019a). It has proven useful in active seismic surveys, mostly VSP, low-frequency strain measurements, and microseismic monitoring. From a practical point of view, DAS is well suited to EGS projects. Fibers can be deployed behind casing and reoccupied for seismic surveying when convenient. As such, any active cased well, either injection or production, can be monitored during operations. More noise is expected in such wells, particularly in the form of tube waves. However, tube wave properties are predictable and have a much lower apparent velocity than the desired signal, and can thus be filtered out during processing. Acquisition using fibers deployed inside active wells is also possible, albeit significantly noisier (Kimura et al., 2019; Uematsu et al., 2019). The fibers can withstand harsh temperature and pressure conditions and can thus record close to or within the reservoir (Zhidong et al., 2019). They can also be left in the subsurface for long periods of time, as they are a purely passive component. For example, at SAFOD (Lellouch et al. 2019b), a fiber was interrogated twelve years after its installation. Nonetheless, especially in high-temperature environments, chemical degradation processes, especially hydrogen darkening, may decrease signal quality with time. Methods have and are currently being developed to address this problem, but sufficient data are not available for a quantitative estimation of the average lifespan of a fiber in EGS conditions. At any time, connecting an interrogator to the fiber at the surface can provide immediate seismic recording, without any effect on field operations.

In a previous study of the FORGE experiment (Lellouch et al., 2020, SRL, in press), we conducted a thorough comparison between DAS and downhole geophones for microseismic monitoring. In the current installation setup, DAS is not performing as well as downhole geophones in terms of event detection and yields a magnitude completeness of $M = -1.4$ versus $M = -1.7$ for a co-deployed geophone array. However, given all the benefits previously described, we think it is worthwhile to estimate its performance in monitoring larger earthquakes as well, in particular when the practical alternative for long-term monitoring is usually surface seismometer arrays.

Using a more advanced workflow, that includes event clustering and template matching, we detect 82 events that are not associated with the stimulation and with visible P and S arrivals within 15 km of the monitoring well between April $23_{rd}$ 00:00 and May $3_{rd}$ 13:00 UTC. In that time period, the regional network reported ten events in the University of Utah Seimograph Stations (UUSS) catalog, out of which only four were not associated with stimulation. We also show that DAS-based event locations, limited to horizontal distance from the array and depth, align with two major source areas of historical seismicity in the region. By examining the seismograms recorded by the regional network at the times of DAS detections, we detected 16 of the 82 events and coarsely locate them in good agreement with the historically active source areas. Finally, we compare seismicity rates during the stimulation period with the background seismicity levels. The number of DAS detections exceeds the expected seismicity levels, indicating the quality and benefits of a downhole DAS-based-detection.

**2 Study area and monitoring arrays**

Figure 1 contains an overview of the FORGE experiment elements involved in this study. We focus on DAS data acquired in the monitoring well. The fiber was installed in a metal tube cemented behind the casing. The well and fiber reach a 985 m depth, crossing into granitic basement at approximately 800 m depth. In order to improve signal-to-noise ratio (SNR), an engineered fiber was used and interrogated by a Silixa Carina system. This form of recording has been shown to improve the optical SNR by 20 dB (Correa et al., 2017). Acquisition has been quasi-continuously active from April $23_{rd}$, 2019 to May $3_{rd}$, 2019. A total of about 40 minutes were not recorded. DAS data have been recorded with a 1-m channel spacing, 10-m gauge length, and 2000 samples per second after a 16-fold internal stacking of the laser sampling rate prior to writing to disk. The output of the DAS interrogator is an optical phase measurement of the strain-rate, which can be converted from radians per second to physical strain-rate (measured in nm/m per second) using a linear conversion.

In addition to the DAS fiber, the monitoring well was equipped with a Schlumberger 12-geophone string, spanning depths of 650 to 980 m. The data recorded by the geophones have also been fully processed and cataloged by Schlumberger, albeit for microseismic events only. No earthquakes from beyond the stimulation zone are present in the downhole geophone catalog, almost certainly due to intended filtration of non-microseismic events. In a previous study (Lellouch et al., 2020, SRL, in press), we compared the performance of downhole geophones and DAS with the downhole array microseismic analysis. However, the geophones did not continuously record for the entire duration of the experiment, and there were significant temporal gaps in acquired data. As the purpose of this study is to evaluate the performance of long-term seismic monitoring with DAS, we did not process the raw geophone data to build an earthquake catalog.

Surface instruments deployed by the University of Utah Seismograph Stations (UUSS) were active before, during, and after the stimulation experiment and are shown as blue triangles in Figure 1. Temporary deployments during the stimulation of additional surface and shallow borehole receivers are not used in this study due to their proximity to anthropogenic noise sources.

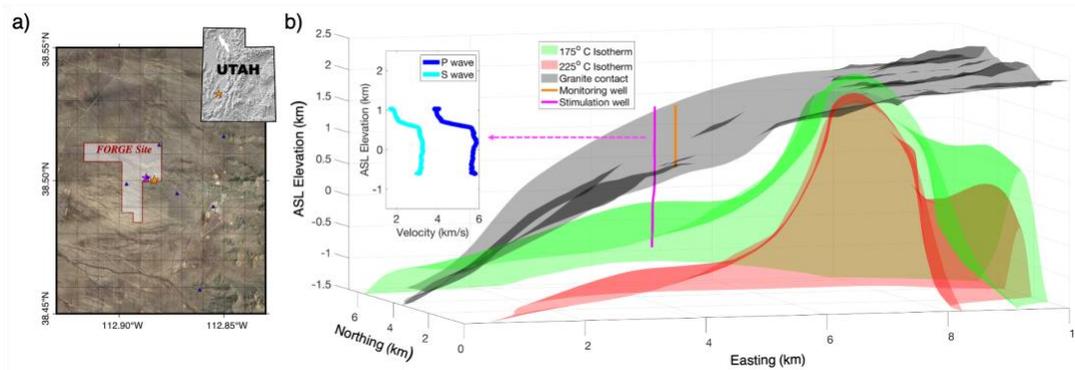

**Figure 1 -** The FORGE site. (a) Aerial view of the FORGE site in Utah. The stimulation well (58-32) is a magenta star, and monitoring well (78-32) is an orange star. Notice the Blundell geothermal plant, located in the Roosevelt Hot Springs to the east of FORGE, visible in the satellite imagery. Surface stations are marked in blue triangles. (b) Local cross-section view of the geological and thermal conditions. We plot two isotherm surfaces (green – $175_0$ C, red - $225_0$ C) and the granite contact (black), obtained from the FORGE database (http://gdr.openei.org/submissions/1107). The stimulation (magenta) and monitoring (orange) wells are overlaid. In the side panel, we show sonic logging of P- and S- wave velocities, acquired in the stimulation well. A sharp velocity contrast is apparent when crossing the granite contact.

The stimulation of well 58-32 (magenta line in Figure 1b) was separated into three phases, each containing nine different stages. The first phase was in an open hole section, and the other two were in areas in which the casing was perforated (Moore et al., 2019). Different stimulation profiles were tested during the various phases. We use the casing pressure measured in the stimulation well throughout this manuscript.

## 3 Earthquake detection and recorded events

3.1 Initial detection method

The workflow we used to detect earthquakes is almost identical to that for microseismic events described in Lellouch et al. (2020). Here we will only briefly recapitulate the main steps. More details can also be found in Lellouch et al. (2019a, 2019b).

1. Data pre-processing – median removal, band-pass filter (5-100 Hz unless mentioned otherwise), removal of noisy channels, and trace-by-trace $L_2$ normalization.

2. Building the P- and S- wave velocity model along the fiber. For the P-wave velocity, this can be done using recordings of the perforation shots, with known locations. For the S-wave velocity, strong microseismic events with good control on location are used.

3. Computing predicted P and S first-arrival times along the array based on the estimated velocity models and angle of incidence, measured in relation to the vertical axis. Events are assumed to reach the bottom of the array first, as planar wavefronts.

4. Finding optimal angle of arrival for continuous data records. Different angles of arrival predict different travel-time curves. For each angle, we align the data along the relative predicted times (no absolute timing), and measure coherency using semblance (Neidell and Taner, 1971).

5. Applying a detection threshold, and aggregating temporally close events to a single detection. This step also yields initial P- and S- arrival time picks.

With this approach, both microseismic events and earthquakes are detected. We filter out all events in the microseismic catalog, based on their detection times. In addition, we manually adjust P- and S- picks to more accurately represent the first arrival times. Such corrections are sometimes required as phase conversions at the granite contact or coda events can have semblance values higher than the first P-wave arrival. This is especially important for weak events, in which only a single phase (either S or P) is detected by the angle scan. By doing so, we also filter out all events in which only a single phase can be manually picked. Following our initial assumptions of planar events, the picks represent the arrival time at the bottom of the DAS array. After all these steps, we obtain a catalog containing 77 events, along with their P and S arrival times.

3.2 Earthquakes recorded by DAS

Examples of DAS records in which earthquakes were detected are shown in Figure 2. They demonstrate the major advantage of DAS – continuous, high resolution spatial and temporal sampling of the seismic wavefield. Even with low SNR, one can see that earthquakes have distinct arrival patterns as function of depth along the fiber. We also observe major differences between the six earthquakes shown in Figure 2. The P-S time difference is quite similar for events (a)-(c), but not for (d) – (f). This indicates that the events are from different locations. The relative amplitude distribution between the P and S phases indicates difference in the source mechanism. For example, in event (b), the S wave is relatively stronger, whereas in event (c), the P wave dominates. While source localization is unfeasible with a single vertical DAS well, due to its azimuthal ambiguity, more complex acquisition geometries could allow for it (Karrenbach and Cole, 2019).

3.3 Matched filtering using DAS records

Given this initial catalogue of event detections, we seek to enrich the number of detections using matched-filtering techniques. Matched-filtering typically has the potential to increase event detections by an order of magnitude or more in conventional seismology studies (Schaff & Waldhauser, 2010). However, the use of this technique to DAS seismology has been relatively limited and used only with surface DAS applications (Li and Zhan, 2018, Yuan et al., 2020). Because of the strong variability between DAS datasets due to instrumentation, optical parameters, and fiber coupling, methodologies that handle multiple spatially-coherent channels are often site-dependent. To ensure the rigorous application of this technique to our downhole dataset, we first re-examine prior rules-of-thumb, such as using thresholds of 0.30 cross-correlation coefficient (CC) in single-station techniques (e.g., Gibbons & Ringdal, 2006; Schaff, 2008). From the catalogue of ~300 detected events (including the previously studied microseismicity from Lellouch et al., 2020), we examine the CC values of all ~50,000 event pairings to get a sense of both event clustering and the statistical distribution of CC values. In this context, CC values are determined by performing N independent cross-correlations for the N channels in the DAS fiber. The N channel correlograms are stacked, using weights based on the average channel-SNR for all events, to enhance stacking performance (Beauce et al 2017; Liu et al., 2020). Based on the statistical distribution of CC values, we ascertain a median absolute deviation (MAD) of ~0.03 CC. Typically, studies have considered detection thresholds of 5-15*MAD (Tang et al., 2010; Huang et al., 2017). We initially consider the 6*MAD value of ~0.20 CC as a reasonable minimum threshold for clustering events into families for matched filtering, which is slightly more sensitive than the often-used CC value of 0.30 for single-channel seismometer detections. This reduces the 77 candidate detections by 3-fold, into 24 distinct families of events that may be used as templates. In addition, to test if this 6*MAD detection threshold is adequate for detection, we apply the matched filtering technique to the entire continuous DAS dataset, using a time-reversed acausal template. By applying the time-reversed template, we assure that the output CC values will be purely random as they do not represent any real correlation between physical events. Therefore, output detections will ascertain the false-positive rate. This test finds that none of the acausal CC values were higher than 0.09.

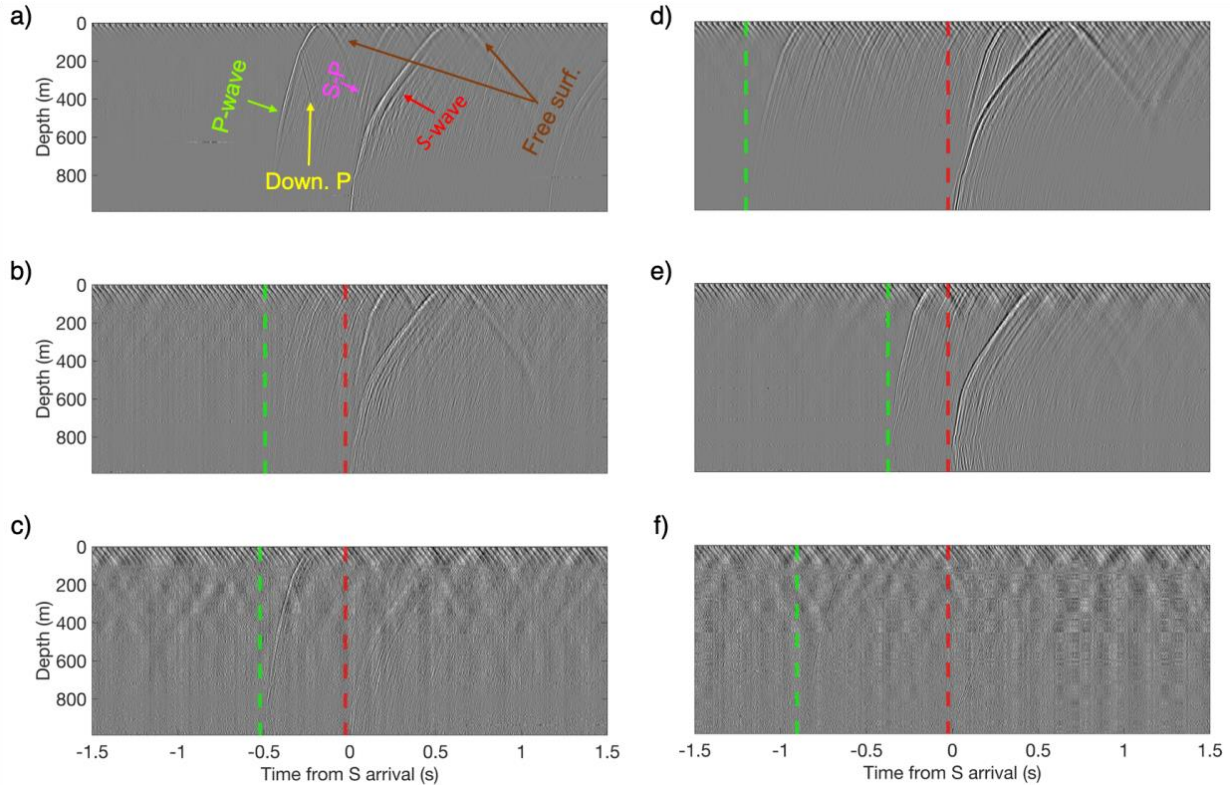

**Figure 2** - Earthquakes recorded by DAS, after pre-processing and filtering between 5 – 100 Hz. All events are centered on the first arrival of the S-phase. Many different phases can be observed, some of which are labeled in (a): direct, upgoing P-waves (green), downgoing P-waves reflected due to impedance contrasts (yellow), direct S-waves (magenta), their conversion to a P-wave at the granite contact (red), and free surface reflections (brown). Coda waves and a secondary event are present but not specifically marked. For events (b) – (f), we plot the picked P (green) and S (red) arrivals at the bottom channel of the array, after a 25 ms shift for display purposes. Events (a) - (c) could originate from the same area, judging by similar the P-S time difference, whereas P-S times for events (d) – (f) are very different from one another and from events (a) – (c). The top (a,d), middle (b,e), and bottom (c,f) lines contain events of high, medium, and low signal-to-noise ratio, accordingly. However, the events are clearly visible in all thanks to their spatio-temporal patterns.

Given a robust matched-filtering approach and detection thresholds, we apply them to the entire continuous DAS dataset, using the 18 out of the 24 families of events as templates. Only families containing more than one event were used in order to increase the SNR of the templates. This template choice was selected to highlight events that are already most active during the DAS recording period and to ensure higher template SNR. To produce a single template from a family of events, we align all events based on their picked S-wave arrival time, and stack them. We then cut the templates at 50 ms before the earliest P-arrival pick in the family to 500 ms after the latest S-arrival pick. Therefore, trailing and leading noise is limited to the necessary minimum. For the detection of new events, we relax the CC cut-off threshold. Instead of 0.20 used for clustering, we use 0.09 CC. While this potentially increases the number of false alarms, it also allows for the detection of weak events. From this analysis we find 32 new events above the cut-off threshold, 16 of which are visibly discernable in the DAS data. None of these events were above 0.20 CC, justifying our choice of a lower threshold. Out of the visible events, 5 have clear P and S arrivals, and we thus include them in our finalized DAS catalog that includes 82 events. This increase in event detections is significantly less accentuated than conventional seismology applications (Schaff, 2008). One potential reason could be related to the primary detection algorithm, that by design, already takes advantage of patterns in the spatially dense and continuous DAS data. Because these were detected using a physical model of wave propagation along the DAS array rather than similarity between events, most of the information gain may already have been accomplished. As a result, the benefits of template matching are, in this case, rather limited. Nonetheless, they ensure that our catalog is complete for events that match the templates.

3.4 Surface array recordings of DAS detections

Four earthquakes outside of the stimulation zone were present in the UUSS earthquake catalog during the experiment (and one within the USGS catalog). We examined the seismograms recorded by the local surface array around the times of the DAS catalog. One of the stations (FORK) is in a shallow borehole. We show the surface seismograms (Figure 3) from the same times as the DAS events shown earlier (Figure 2). Based on the timings of DAS events, we visually inspect the data (Figure 3) using the Antelope software package. Phase arrivals are manually picked and events located (Pavlis et al., 2004) using the local velocity model, which is a combination between DAS-derived models and sonic logging, and depth constraints from the DAS recordings. We use extrapolated constant velocities below the bottom of the logged area. Such a simplified model carries limitations and is expected to affect location results. For these events, we could manually detect and locate the earthquakes using at least two surface stations and four P/S picks.  Events were predominantly located using three stations, with the most clearly defined hypocenters recorded on up to seven stations.  These relocated events, discussed in a later section, confirm that at least 16 of the DAS detections were related to the ongoing seismicity underlying the Mineral Mountains.  We find it likely that all of the DAS-detected seismicity was also related to these swarms (details are elaborated on in a later section).  We note that these events were likely below the standard completeness magnitude of the UUSS.  The feebleness of these events likely also impacts their reported spatial distribution, as events were only visible at the western most edges of the source region that is most adequately covered by local seismometers.

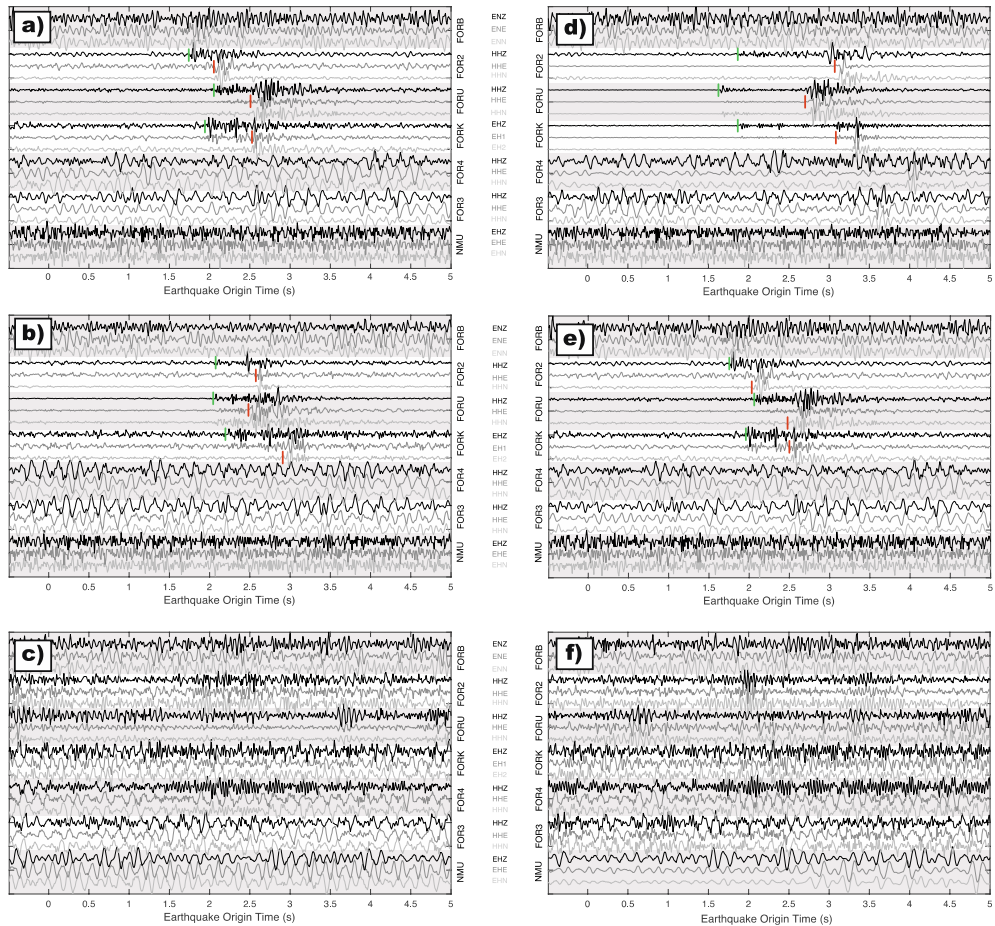

**Figure 3** – Surface array recording of the earthquakes shown in Figure 2. Events (a) – (d) are visible in several stations, despite very low signal-to-noise ratio. P (green) and S (red) picks are plotted where applicable. The vertical component is not sufficient for S-wave picking in the majority of the cases. Events (e) and (f) are below the noise level and no events can be detected, despite being clear in the DAS recording.

As we show later, the majority of the surface stations are closer to the earthquakes than the monitoring well. The events are unequivocally clearer in the DAS records. It is worth mentioning that the surface seismograms are unusually noisy in the bandwidth of local earthquakes, because of wind and anthropogenic activity (McNamara & Buland, 2004). As can be seen from the DAS records, surface noise at depths of 100 m and more is negligible even when compared to the weaker earthquakes. This result is corroborated by prior studies on the impact of emplacement depth on station performance (Hutt et al., 2017). As such, DAS records enjoy the benefits of a much quieter environment for the majority of the channels.

3.5 Summary of detected events and stimulation activity

Figure 4 summarizes both DAS and surface detections during the stimulation period. We use a semblance measure to estimate the certainty of the detections using the DAS array. It represents the maximal semblance (on a scale of 0 to 1) value for either the P or S phase obtained during the angle scan. We use it as a proxy for event clarity and certainty in the DAS data. Figure 4 shows several temporal clusters. In addition, it shows that the surface array is clearly biased towards the detection of events with higher DAS semblance, which is rather unsurprising. The average semblance value for all DAS events is 0.17, whereas it is 0.31 for the events identified by the surface array.

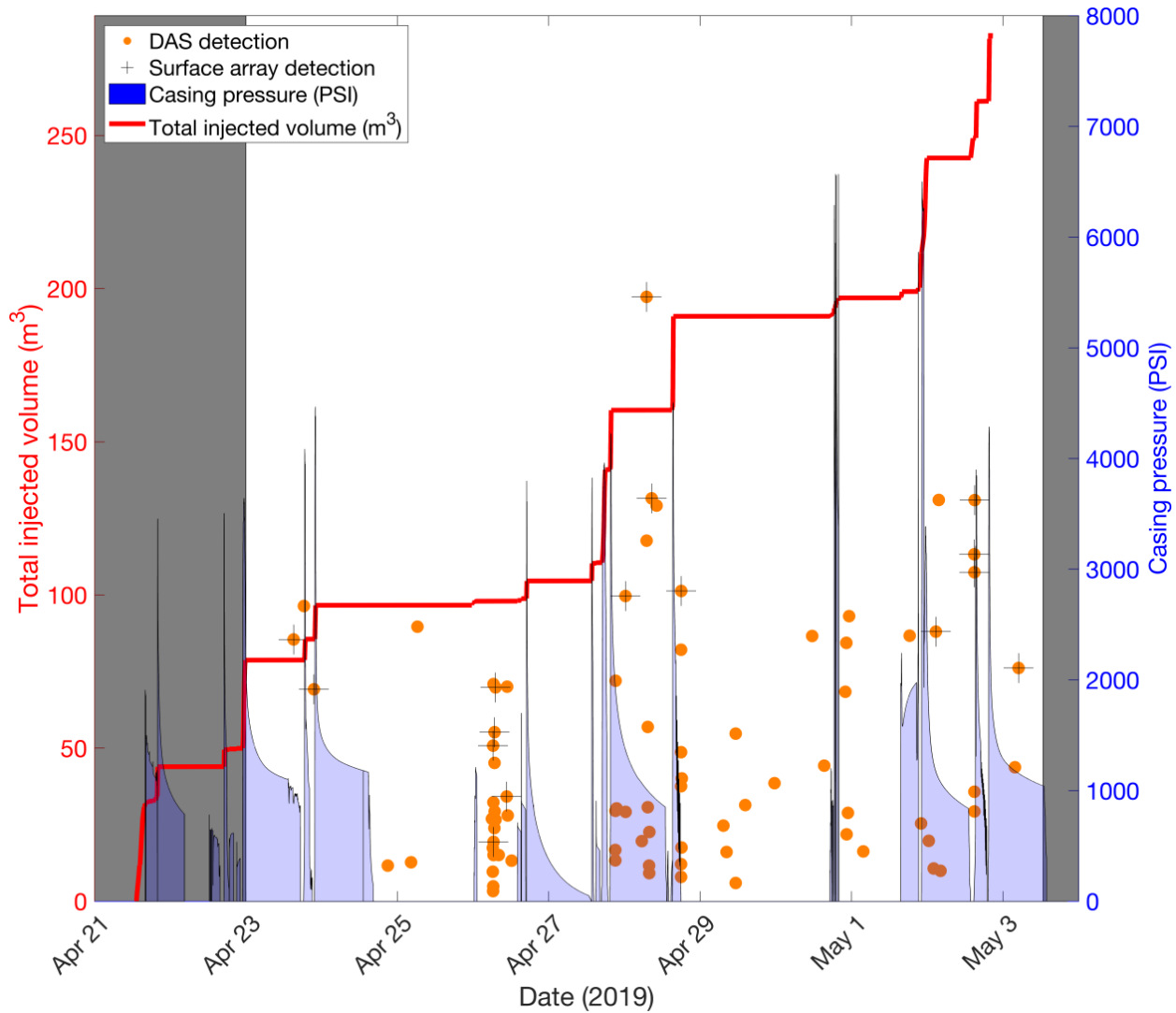

**Figure 4** - Summary of DAS and surface event detection and stimulation operations. The DAS acquisition was not continuous in the grey intervals and we thus did not process the data from that period. DAS detections are plotted in orange circles as a function of time and maximal semblance value obtained from the angle-scan procedure. Semblance values (0-1) are scaled to the vertical dimension of the figure. Several clusters are visible by their temporal density. Events that are also visible on the surface array are additionally

denoted by black crosses. The casing pressure is in blue and shows the three different stimulation intervals. The total volume of injected fluid is in red, indicating that overall, a very limited volume of fluid was injected. The flowback rate is not shown.

## 4 Event analysis

### 4.1 Event location using DAS

DAS records can be used to estimate the distance from the array and depth of the source. We have previously discussed measurement of P- and S- arrival times at the bottom of the array (Figure 2). With the additional information on the P- and S- angles of arrival provided by the semblance scan, we can estimate the focal depth and horizontal distance from the well. The azimuthal information is, as always with a single vertical DAS array, lost. A practical issue arises when the estimated angles of arrival differ between the P and S phases. In the 26 cases where both P and S phases are above the semblance threshold, the mean difference between the P and S estimated angle is about 1.4 degrees, with a standard deviation of 9.5 degrees. However, for the remaining 56 events, either P or S are below the semblance threshold, and the angle estimation in that case is unstable. We opt for choosing the phase that yielded a higher semblance and use the respective angle.

Our location approach is based on two major assumptions. First, for computational feasibility of the detection method, we assume that any earthquake reaches the bottom of the array first, as a planar wavefront. Otherwise, there is not a single angle of arrival, and an infinite number of different travel-times curves can be possible. While it is possible to visually separate events that do not abide to this criterion, i.e. do not reach the bottom of the array first, and process them differently, this is not the focus of this study. The second assumption is that once the angle of arrival is estimated, at the bottom of array, the event location is computed using a constant velocity assumption. In other words, it uses straight rays with an initial vertical slowness estimated from the angle scan, and positions the events along that ray using the S-P time difference. It is reasonable to assume straight-ray propagation in the granitic basement up to a depth of ~1 km ASL as there is very little variation with depth (see sonic logs in Figure 1). Assuming that the velocity of the basement will eventually increase with depth, bending ray paths can be expected. Therefore, our location method is likely to overestimate the depth of events. It is expected to deteriorate with distance, as the cumulative effect of ray bending will increase. Another effect of this assumption is that no event can be located above the bottom of the well, as the maximal possible angle of arrival is $90^0$ and straight rays are used.

A better way to conduct the location procedure would be using ray tracing with a takeoff angle estimated using the angle scan procedure. The ray would be propagated through a velocity structure, and the event would be positioned according to the total distance traveled along that ray. For this method, knowledge of the velocity structure below the maximal logged depth would be required, possibly obtainable from surface seismic surveys (Moore et al., 2019). However, this is not the focus of this study, and DAS location has more pressing limitations such as the azimuthal ambiguity.

Figure 5 summarizes the DAS location and compares them to those of historical seismicity in the region. While a single vertical well suffers from azimuthal ambiguity, the projection of historical locations onto the distance-depth plane reveals strong correlation with the DAS locations. As mentioned earlier, the constant velocity model used is the most probable reason for the depth overestimation of events in the distant source area. In addition, due to our assumptions, no event can be estimated above the bottom of the well, and there is a clustering of events at that depth.

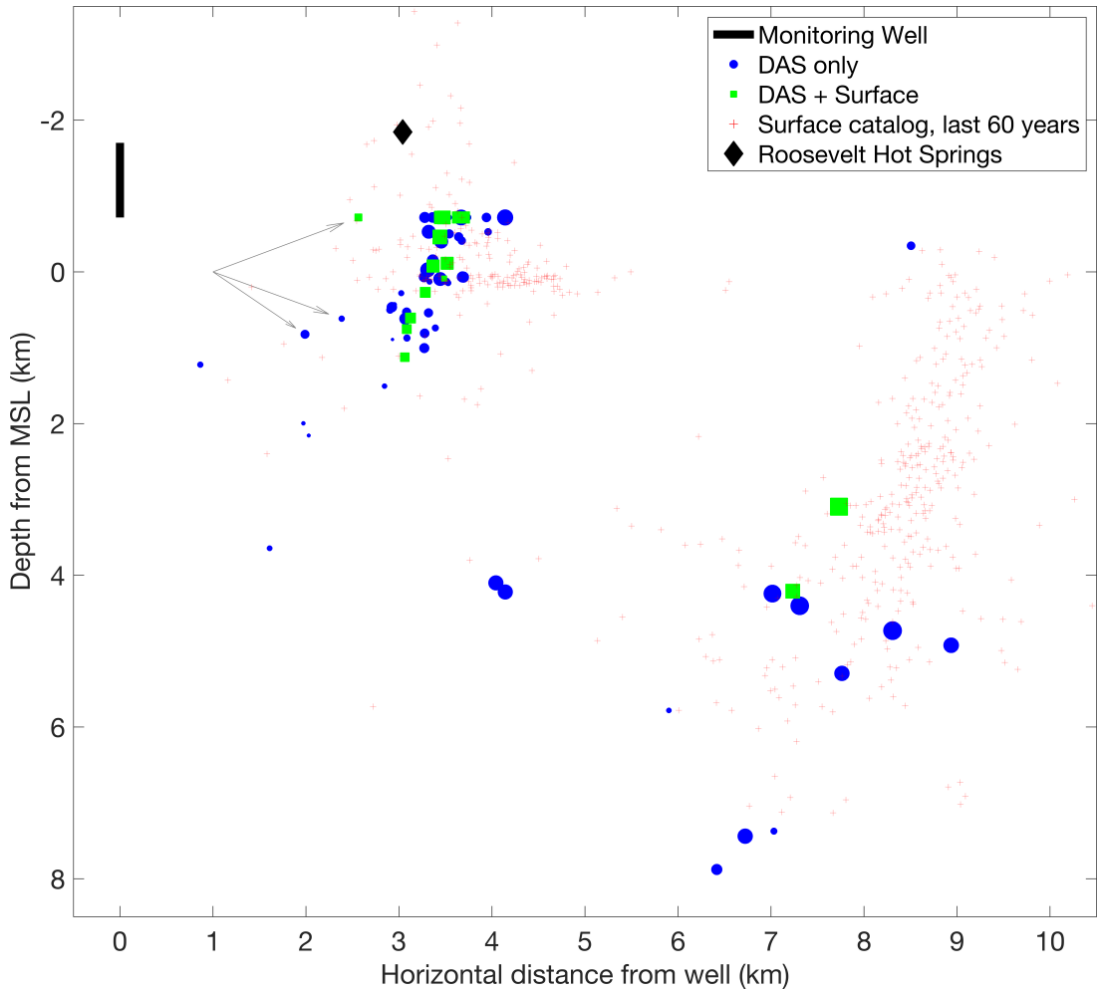

**Figure 5** - DAS-based event location. Monitoring well is in black. Located events are plotted as a function of horizontal distance from the well and depth. Blue circles indicate events detected by DAS only, and green squares are for events also visible in the surface array. For both types, the size of the marker is proportional to the semblance, and hence certainty in the location. Locations of historical seismicity, taken from the University of Utah Seismographic Stations earthquake catalog in the past 60 years, are marked as faint red crosses and do not contain magnitude information. Three events that we visually classify as refracting along the granite are pointed at with grey arrows. The green one is shown in Figure 6. In this 2-D projection, DAS locations appear to agree with locations of historical seismicity, except several uncertain points.

4.2 A wrongly located shallow earthquake

Figure 6 depicts an interesting earthquake that does not abide to our primary assumption about plane-wave arrivals to the bottom of the array. In contrary to the vast majority of events, the first arrival occurs at a depth of ~840 m, close to the granite contact estimated at ~800 m. This is indicative of an event originating above the granite contact and refracting off it. The angle scan and subsequent location place him exactly as the depth of the well, as for both P and S waves its angle is estimated at the possible maximum of $90_o$. Its depth estimation is wrong; it could be located anywhere above the granite contact, whose depth is laterally changing (Figure 1). As the contact is shallower moving towards the east, we suspect that its upper depth boundary is shallower by several hundred meters. Two similar events marked in Figure 5 have lower SNR and their angle of arrival is thus wrongly estimated, but they also do not abide to the basic assumption. We want to emphasize that we are able to conduct this detailed analysis thanks to the spatial resolution and continuity of DAS.

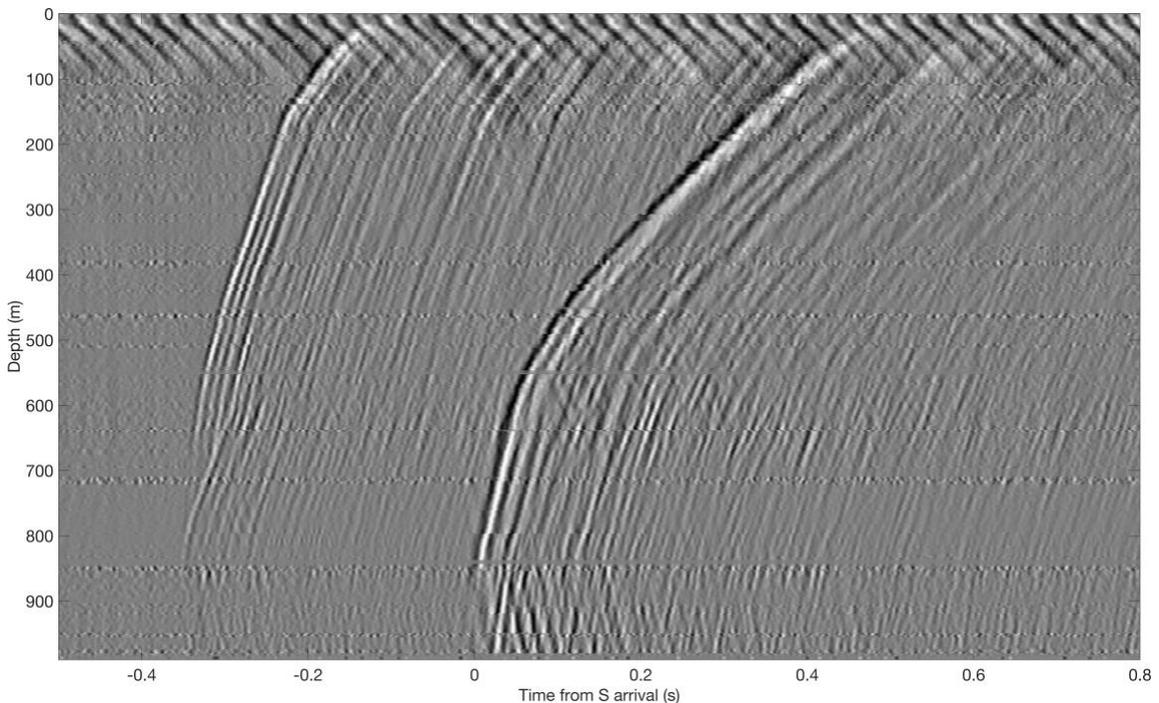

**Figure 6** - DAS record of an event refracting off the granite contact. The first arrival is not observed at the bottom of the array as for the majority of events. Instead, it appears at about 40 m below the granite contact, estimated at around 800 m depth. This property holds for both P- and S- arrivals. The recorded signal is much weaker in channels deeper than 840 m, as can expected from a refracting event. This event was previously shown in Figure 2d, and it strongly differs from earthquakes originating below the granite contact.

4.3 Location validation using surface stations

We also locate the 16 events that could be detected in the surface array to validate the DAS locations. We emphasize that these are previously uncatalogued events, with low SNR and usually very few picks. Therefore, the locations are not expected to be precise. Figure 7 shows that the majority of the events are indeed originating from the source area underlying Mineral Mountains closest to the monitoring well. The surface derived event locations (within the local array footprint) do show some degree of accuracy: events coincide with a previously known activity, and some microseismic events related to stimulation (not shown in the figure) were also recognized and located closely to the FORGE stimulation well. However, outside of the footprint of the local surface array, two DAS-derived events suspected to be related to the further/eastern swarm show a west bias in their surface derived location. This west-bias in locations could be related to the inaccuracies in our velocity model, especially with a positive velocity gradient within the granite, discussed earlier. Given such a gradient, S-P differences can translate to longer distances, which despite ray-bending could still displace events further east. We note that these velocity model biases are in addition to the aforementioned sparsity of arrival picks and station geometry.

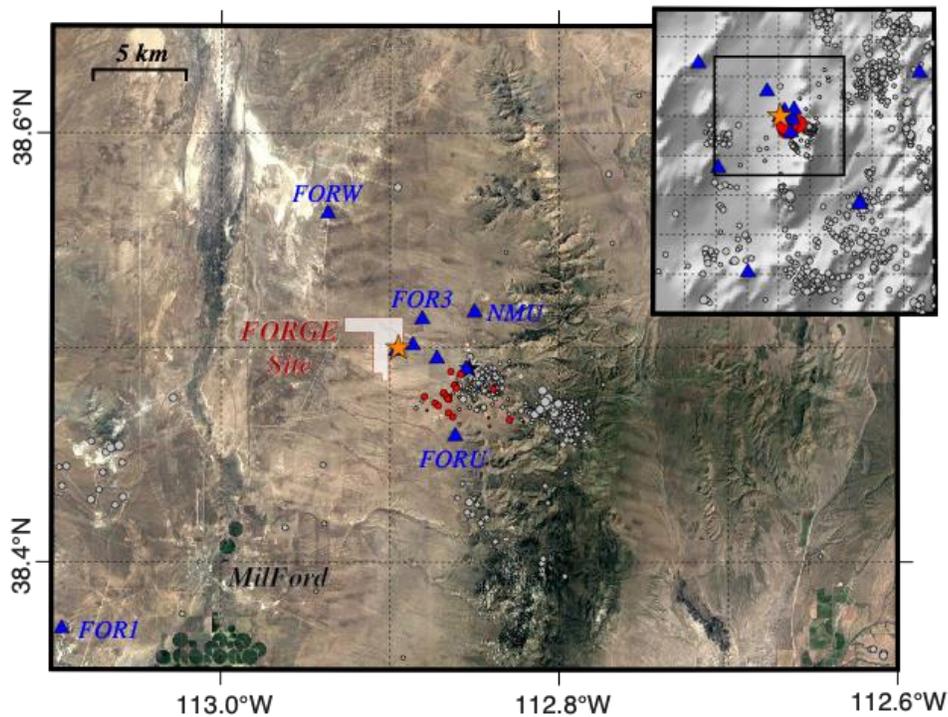

**Figure 7** - Event locations using the surface array. The FORGE site is marked in a red polygon, with the monitoring well as an orange star and the Blundell Geothermal Plant as a black star. Surface stations are in blue triangles. Historical seismicity in the region is displayed as gray circles, with size proportional to their magnitude. Two distinct clusters can be observed underlying Mineral Mountains. Our surface-based locations indicate that the detected events are generally originating from the nearer cluster. The location quality

is poor due to low SNR and number of stations that record the events. Two events that appear in the DAS data as more distant are separated from that cluster.

4.3 Clustering DAS events

Clustering approaches (Kaufman & Rousseeuw, 2009) have often been applied to discern similarity between earthquake waveforms. In Figure 8, we show a final application of the clustering algorithm repeated on the full DAS catalog containing all 82 events. This approach provides a simple means to examine and quality control potential groupings of events. Our similarity matrix is grouped into clusters via an agglomerative hierarchical cluster linkage algorithm using an average distance metric and a CC threshold of ~0.06 chosen near the MAD that visually highlights groupings. We note that this choice of CC threshold is more lenient than the one used for the template matching, and thus produces fewer families. This average metric was chosen to emphasize overall groupings of clusters, rather than nearest-neighbor type groupings. We then compare this CC-derived clustering to the DAS-derived radial distances and event detection timings. Clear delineations of clusters are noted between these datasets (Figure 8). For example, ongoing swarm activity is noted at 3-4 km distance, with punctuated bursts of relatively independent clusters. In general, clusters tend to be at distinct distances (e.g., C2, C8, & C7). This clustering approach provides some corroborating evidence to the DAS derived results. Although, we also note that the current dataset produces distinct clusters that are not readily discernible (e.g., C2 vs C1, C5, & C4). Likely, additional information on earthquake focal mechanisms would be useful to better discern the subtleties of the inter-cluster groupings.

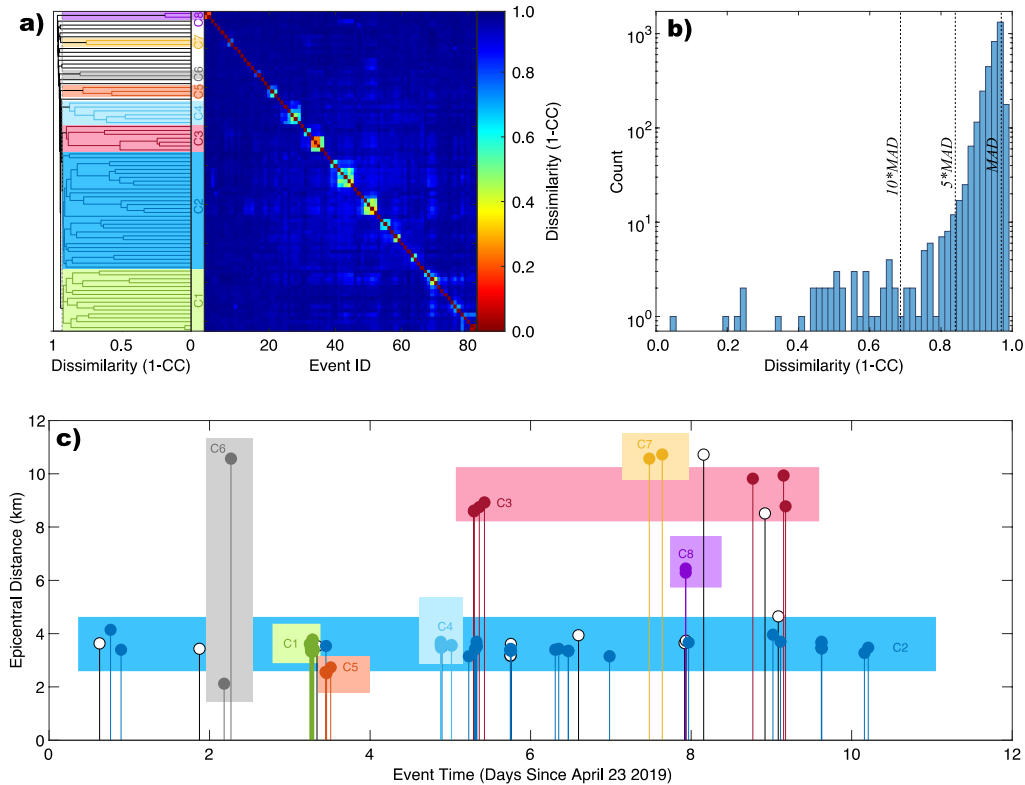

**Figure 8** - Agglomerative hierarchical clustering of DAS data. a) The similarity matrix (center box) displays the CC values of all 82 event pairs (see scale bar). The ordering of events has been sorted, via a clustering approach shown in the associated dendrogram.

4.4 DAS magnitude estimation

We use the DAS records to estimate magnitudes as well. Our approach is only approximate, as it is based on empirical relations, uses only a single component of the measurement, and does not account for fiber response. We describe the methodology and show (Lellouch et al., 2020, SRL, in press) that for microseismic events, magnitudes are in good agreement with a downhole-geophone-based catalog. In this study, we use the same approach, based on measuring cumulative strain in the DAS channels, but with events filtered between 10 – 100 Hz. After applying this methodology, we obtain a magnitude distribution that we compare to the last five years of the surface catalog (Figure 9). It is difficult to conduct a meaningful statistical analysis based on only 82 DAS events, in which the magnitude estimation is only approximate. However, this comparison unequivocally shows that the DAS array is more sensitive than the surface one. While it is difficult to accurately quantify the difference between them, we estimate the DAS catalog is between 0.5 to 1.0 magnitude units more complete. This is corroborated by the fact that from the 82 events detected by the DAS array, only one was in the surface catalog. The b-value obtained for the DAS array is not statistically significant enough to be treated reliably, but the value estimated (b~1) using the last five years of the surface catalog is probably indicative of the regional seismicity. Details of the magnitude distribution of events is elaborated on in the following section.

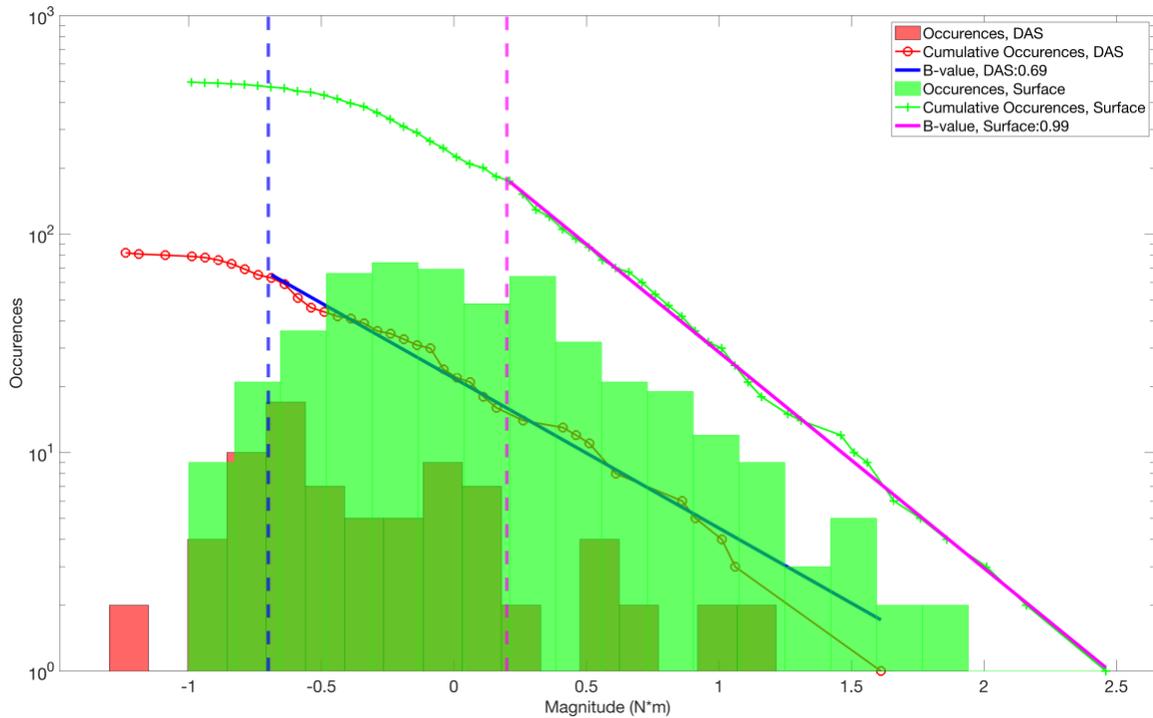

**Figure 9** - Magnitude estimation using DAS for the 10.5 day FORGE experiment compared to the 5.5-years USGS catalog based on surface stations. DAS event occurrences are plotted as a red histogram, their cumulative distribution in red circled line, and the maximum likelihood fit to that line in blue. We estimate the DAS FORGE catalog to be complete above M= -0.7 (dashed vertical blue line) and use those magnitudes to estimate a b-value of 0.69, however due to the brief recording duration and low number of events, these values are unreliable. The much longer USGS catalog is shown with a green histogram of occurrences, green crossed lines for the cumulative distribution, and the maximum likelihood fit to that line in magenta. The b-value is very close to 1, and the catalog is complete at M=0.2 (dashed vertical magenta line) and above.

**5 Natural or FORGE-induced seismicity?**

Enhanced geothermal systems and the process of stimulation by hydraulic fracturing are known to induce or trigger earthquakes (Majer et al., 2007; Grigoli et al., 2018; Schultz et al 2020). Due to large number of events detected by DAS, we examine the possibility that the events detected by DAS outside of the stimulation volume were induced by stimulation of the 58-32 well in the FORGE project. It is important to note that the total amount of injected fluid was very low (<300 m3), and thus it is highly unlikely that they are caused by induced seismicity.

To begin answering this question, we must first understand the regional seismotectonics. In general, central Utah has a tectonically complex concentration of seismicity in the Intermountain Seismic Belt; the transition area between the northwestern Colorado Plateau and the eastern

Great Basin of the Basin and Range (Arabasz & Julander, 1986). Natural seismicity in Utah exhibits both strike-slip and normal faulting with tens of M3+ events occurring per year (Arabasz et al., 2016). In the broader context, induced earthquake cases have been noted in association with both coal mining (Arabasz et al., 2005; Pechmann et al., 2008) and subsurface injection operations (Brown & Liu, 2015). Focusing slightly closer to the FORGE project, in the Marysvale volcanic field, swarm activity dominates the seismic activity, is typically associated with events in the upper crust and is thought to be related to either hydrothermal or volcanic activity (Arabasz et al., 2007). The largest historical event nearby to our study area was a M 4.1 south of Milford in 1908, with more recent events indicating steeply dipping strike-slip faults (Whidden & Pankow, 2012; Potter, 2017).

To the southeast of the FORGE project site (4 & 8 km), two concentrations of seismicity underlying the Mineral Mountains are apparent in the UUSS earthquake catalog (Figure 8). Corroborating the swarm-volcanism association, low velocity bodies under this mountain have been interpreted as partial melt related to volcanic activity (Robinson & Iyer, 1981). These swarms were described in studies related to FORGE, characterizing the baseline rates of earthquakes in the immediate area (Pankow et al., 2019; 2020). Most notably, a study of these Mineral Mountains swarms was conducted prior to the development of the Blundell power plant exploiting the Roosevelt Hot Springs geothermal system (Zandt et al., 1982). This study found more than 1000 small magnitude ($M_L < 1.5$) normal faulting events, trending along (but offset from) the previously mapped Negro Mag Fault. Zandt et al., (1982) concluded that these swarms were likely related to natural hydrothermal fluid-flow processes and not related to geothermal pumping activities. Despite this, they acknowledged that these swarms indicated a potential susceptibility of nearby geothermal systems to induced seismicity.

To determine if a sequence of earthquakes might have been induced or not, a set of criteria were established (Davis & Frohlich, 1993; Verdon et al., 2019). The first of these criteria measure the statistical significance of changing earthquake rates. To begin a rudimentary examination of the rates of seismicity, we analyze ~5.5 years' worth (Jan 2015 – May 2020) of UUSS-recorded swarm seismicity near Mineral Mountains (Figure 9). Rates are established though a maximum likelihood parameter estimation of the frequency magnitude distribution of events (Gutenberg & Richter, 1945; Marzocchi & Sandri, 2009; Schultz et al., 2018), where the magnitude of completeness is estimated as the value that maximizes the goodness-of-fit (Woessner & Wiemer, 2005). Based on this, we estimate that the rate of M0+ events in this area is on the order of 50 events per year – and that around six events of M -0.6 or greater would be expected for the 10.5 days of DAS recording. Our visually confirmed DAS events with M>-0.6 during this time is higher (52), but on the order we would expect from natural rates. Obfuscating this result, swarm seismicity has the propensity to naturally change earthquake rates by orders of magnitude (e.g., Klein et al., 1977; Farrell et al., 2009; Crone et al., 2010). In the hydrothermal context, this is often thought to be related to episodic slip where fault valving processes allow for a transient migration of fluid along faults (Sibson, 2020). In particular, swarms underlying Mineral Mountains have been documented to naturally change rates by an order of magnitude over a period of weeks (Zandt et al., 1982). Based on this rationale, we conclude that we are unable to discern any systematic differences from natural variabilities. More enriched catalogues, over a greater period of time, would be required to better discern the potential for triggered or induced seismicity.

## 6 Discussion

This study shows that a vertical downhole DAS array can retrieve microearthquakes at a range of up to 10 km from the well. A single DAS well shows a clear benefit over the available surface array. To obtain these results, DAS data undergo array processing methods that take advantage of their dense spatial continuity. We estimate an improvement of 0.5 to 1.0 in the magnitude completeness, going well into the negative magnitude range, despite the fact that the well is located 3-4 km from the nearest seismically active region. We therefore believe that for M<0 seismic events, the usage of downhole DAS should be further explored. This reinforces our previous study's results at the SAFOD borehole, which also found that vertical fiber DAS-based earthquake detection capabilities are at least as good as a surface network (Lellouch et al., 2019a). However, it is difficult to make sustained claims on the potential and capabilities of DAS for earthquake monitoring, whether natural or induced, based on 10.5 days of recording. The statistical certainty can only be achieved through much longer experiments.

The spatial and temporal continuity of DAS records allows for a much deeper physical understanding of recorded events. Studies regarding coda waves, attenuation, scattering, and so forth will inherently benefit from spatially continuous observation of the wavefield. The direct observation of an event refracting off the granite contact is yet another example of the wealth of information DAS can provide. However, event location, magnitude and focal mechanism are often more important. In terms of location, a single vertical DAS well will always be limited due to its azimuthal ambiguity. We have shown that derived distances and depths agree with known seismic source areas. A 3-D location may be obtained by using additional wells. Two wells would yield two possible mirrored locations for each event, and three would be sufficient for a unique location. Alternatively, surface or shallow DAS can be deployed at a much lower cost. However, as we have seen, data quality in the shallow section may be problematic, especially in the presence of anthropogenic noise, and its advantage over standard surface receivers remains to be proven. Deep DAS channels are much more immune to noise, which is especially important close to anthropogenic activities. If a horizontal DAS array could be deployed at a depth of ~100 m and more, it may prove extremely useful.

Therefore, our view is that DAS can complement existing local networks, even in the form of a single well. All the DAS events we analyze are from a time period in which there were four catalogued events. We found 82 using DAS, and 16 of those were even visible on the local surface network when looking for the events based on the DAS catalog; only four of these events was found in the routine UUSS catalog. The number of events detected during the FORGE experiment is much higher than the background rate of events in the surface catalog, acquired during significantly quieter periods. The potential enrichment of the catalog and the ability to analyze each event with unprecedented resolution underscores the value of borehole DAS data for monitoring microearthquakes.

Magnitude estimation and focal mechanism estimation using DAS still remain an open question, despite our simple yet useful approximation of the magnitudes. A true moment magnitude can only be obtained if the source mechanism is fully described. Conducting this with single-component DAS records is challenging, especially as the fiber response is hard to quantify in the field (Lindsey et al., 2020). More complex acquisition geometries have shown promise for DAS-based focal mechanism estimation (Karrenbach et al., 2019), but research is still in its early stages.

Despite these limitations, our joint analysis of DAS and local surface stations yielded an improved understanding of the seismicity during the period of the FORGE stimulation experiment. While seismicity outside of the stimulation volume was more active than the long-term rate for area, it is not directly related to the FORGE injections. Changes can be attributed to the natural variability of swarm activity, with the closest active area potentially affected by the Blundell geothermal plant.

## 7 Conclusions

DAS holds many operational benefits for monitoring Enhanced Geothermal Systems. It is resistant to heat and pressure with fiber adaptations to more extreme environments, can operate for extended periods of times, and can be acquired in an active well. In this study, we show the potential of a downhole DAS array in an EGS setting. Using an array processing workflow, DAS detects 82 events from outside of the stimulation volume in a 10.5-day period in which only four events were found in the UUSS catalog. 16 of these events can be visually identified in the local surface array, albeit with lower signal-to-noise ratio. We estimate the magnitude completeness obtained using DAS to be better by $M=0.5$ to $M=1.0$ than the surface network. While locations obtained from a single DAS well suffer from azimuthal symmetry, more elaborate geometries can lead to accurate location estimations, and better help constrain the magnitude and focal mechanism. For now, we are only able to use a simplistic approach to magnitude estimation. A joint analysis of DAS and surface data reveals clear increase in seismicity at a distance of 3 to 10 km from the treatment well during the FORGE stimulation experiment. It cannot be attributed to the stimulation and is more likely due to the natural variability of earthquake swarm seismicity.

**Acknowledgments, Samples, and Data**

None of the authors have any conflict of interests. Both geophone and fiber data have been made openly accessible by the FORGE project and scripts for downloading data are available at the US DOE Geothermal Data Repository, along with the geophone catalog. Seismometer data for stations in the University of Utah Seismographic Stations were accessed through the Incorporated Research Institutions for Seismology (https://ds.iris.edu/mda/UU/). The catalog of DAS events can be accessed on https://github.com/ariellellouch/FORGE or 10.5281/zenodo.3909840 (Lellouch, 2020). AL was partially supported by the Israeli Ministry of Energy under the program for postdoctoral scholarships in leading universities. RS was supported by the Stanford Center for Induced and Triggered Seismicity. NL was supported by the George Thompson Postdoctoral Fellowship.

We are much obliged to Kristine Pankow, who introduced us to the FORGE experiment and supported us in accessing the data and auxiliary information. Maria Mesimeri gave us very useful comments throughout fruitful discussions.